\documentstyle[dina4,epsfig,12pt]{article}
\newcommand{\be}{\begin{equation}}   
\newcommand{\ee}{\end{equation}}   
\newcommand{\bea}{\begin{eqnarray}}   
\newcommand{\eea}{\end{eqnarray}}

\newcommand{\non}{\nonumber}

\begin{document}

\title{Probing nuclear expansion dynamics with $\pi^-$/$\pi^+$-spectra
\protect\footnote{Supported by BMBF and GSI Darmstadt}\hspace{3mm}\footnote{part of the
PhD thesis of S. Teis}}
\author{S. Teis, W. Cassing, M. Effenberger, A. Hombach, U. Mosel and 
	  Gy. Wolf \protect\footnote{Supported by Hungarian OTKA Funds T16594,T16249}\\
	Institut f\"{u}r Theoretische Physik, Universit\"{a}t Giessen \\
	D-35392 Giessen, Germany}
\maketitle
\begin{abstract}
We study the dynamics of charged pions in the nuclear medium via the ratio 
of differential $\pi^-$- and $\pi^+$-spectra in a coupled-channel BUU (CBUU)
approach. The relative energy shift of the charged pions is found to 
correlate with the pion freeze-out time in nucleus-nucleus collisions 
as well as with the impact parameter of the heavy-ion reaction. Furthermore, 
the long-range Coulomb force provides a 'clock' for the expansion of the 
hot nuclear system. Detailed comparisons with experimental data for $Au + Au$ at 1 GeV/A and $Ni + Ni$
at 2.0 GeV/A are presented.
\end{abstract}

\vspace{2cm}
 
\newpage
\begin{section}{Introduction}
The meson-baryon dynamics in relativistic nucleus-nucleus collisions is
of particular interest since one expects to learn about the hadron properties
and abundancies in hot and dense nuclear matter. Whereas dileptons from
$\rho^0$ decays or $\pi^+\pi^-$ annihilation provide information on the
timescale of $1.3$ fm/c in the high density phase 
\cite{1,2,3,cassing95,wolf90,wolf93}, pion
correlations can be used to learn about the freeze-out volume of the pions
during the expansion phase. Whereas the latter method has been extensively
explored in nucleus-nucleus collisions at AGS and SPS energies \cite{QM95},
pion correlation measurements are less promising at SIS energies due to the
much lower pion multiplicities achieved at 1 - 2 GeV/A. Alternatively, 
one might
here use the kinematic shift of $\pi^-$ to $\pi^+$ spectra due to the long
range Coulomb force \cite{Muentz} in order to learn 
about the baryon expansion dynamics at 
the production time of those pions that no longer interact with the baryons
and are identified experimentally \cite{Bass,li}.

In this paper we investigate more systematically the pion dynamics in the
time dependent Coulomb field generated by all charged hadrons in
nucleus-nucleus collisions from 1 - 2 GeV/A for systems at SIS.
Our analysis is carried out within the coupled-channel BUU (CBBU) 
transport approach \cite{teis96}
which is briefly described in Section 2. In Section 3 we analyse the
correlation between Coulomb effects and the pion production time as 
a function of the
pion momentum in the cms and study the sensitivity of the $\pi^-/ \pi^+$
spectra to the reaction geometry of the heavy-ion collision. Detailed 
comparisons with experimental data for the $\pi^-/\pi^+$ ratio are, furthermore, presented for
$Au + Au$ at 1 GeV/A and $Ni + Ni$ at 2.0 GeV/A. Section 4, finally, concludes
our study with a summary and a discussion of open problems.
\end{section}
\begin{section}{The CBUU-Model}
\begin{subsection}{Basic equations}
For our present study we use the CBUU-transport-model \cite{teis96} to 
describe the time evolution of relativistic heavy-ion collisions. 
In this approach, apart from the nucleon and the $\Delta(1232)$, all
nucleon resonances up to masses of $1.95$ $GeV/c^2$ are taken into account as 
well as the mesons $\pi$, $\eta$, $\rho$ and $\sigma$. 
While the $\pi$-, $\eta$- and $\rho$-mesons correspond to physical 
particles, the $\sigma$-meson is introduced to describe correlated pion-pairs 
with total spin $J = 0$. For the 
baryons as well as for the mesons all isospin degrees of freedom are 
treated explicitly. The hadrons included in our model obey a set of coupled  
transport equations for their one-body phase-space distributions  
$f_i(\vec{r},\vec{p},t)$ \cite{bertsch88,cassing88,cassing90,kweber93}:
\bea 
\label{buueq}
& & \frac{\partial f_{1}(\vec{r},\,\vec{p_1},\,t)}{\partial t} + \left\{ \frac{\vec{p_1}}{E_1} + 
\frac{m_1^*(\vec{r},\vec{p_1})}{E_1}\, \vec{\nabla}_p\, U_1(\vec{r},\, \vec{p_1}) \right\} 
\, \vec{\nabla}_r f_{1}(\vec{r},\,\vec{p_1},\,t)  \nonumber  \\
& & \hspace{2.5cm} + \left\{ -\frac{m_1^*(\vec{r},\vec{p_1})}{E_1}\, \vec{\nabla}_r  
U(\vec{r},\, \vec{p_1}) -q_1 \vec{\nabla}_r 
V_C(\vec{r})\right\}\, \vec{\nabla}_p 
f_{1}(\vec{r},\,\vec{p_1},\,t) \non \\
& &   =  \sum_{2,3,4}
\frac{g}{(2\pi)^3} \, 
\int  d^3 p_2 \, \int  d^3p_3 \, \int  d \Omega_4 \, 
\delta^3 \left(\vec{p}_1 + \vec{p}_2 
- \vec{p}_3 - \vec{p}_4 \right) \nonumber \\
& &\hspace{2cm} \times(
v_{34}\, \frac{d\sigma_{34 \rightarrow 12}}{d\Omega}\, 
f_3(\vec{r},\,\vec{p}_3,\,t)\, 
f_4(\vec{r},\,\vec{p}_4,\,t)\, \bar{f}_1(\vec{r},\,\vec{p}_1,\,t)\,
\bar{f}_2(\vec{r},\,\vec{p}_2,\,t)) \nonumber \\  
& & \hspace{2cm} -  \, \, \, 
v_{12}\, \frac{d\sigma_{12 \rightarrow 34}}{d\Omega}\, 
f_1(\vec{r},\,\vec{p}_1,\,t)\, 
f_2(\vec{r},\,\vec{p}_2,\,t)\, \bar{f}_3(\vec{r},\,\vec{p}_3,\,t)\,
\bar{f}_4(\vec{r},\,\vec{p}_4,\,t)).
\eea
The l.h.s. of eq. (\ref{buueq}) represents the relativistic Vlasov-equation for 
hadrons with charge $q_1$ moving in a scalar 
momentum-dependent field $U_i(\vec{r}, \vec{p_i})$ \cite{teis96} as well as 
in the electromagnetic potential $V_C(\vec{r})$, where $\vec{r}$ and $\vec{p_i}$ 
stand for the spatial and momentum 
coordinates of the hadrons, respectively. The effective mass 
$m_i^*(\vec{r}, \, \vec{p_i})$ in 
eq. (\ref{buueq}) includes the restmass $m_i^0$ of hadron $i$
as well as a scalar mean-field potential 
$U_i(\vec{r},\, \vec{p_i})$, i.e. 
\be
\label{effmass}
m_i^*(\vec{r}, \, \vec{p_i})  = m_i^0 + U_i(\vec{r},\, \vec{p_i}).
\ee
Since in our model the mesons will be propagated as free particles with respect 
to the nuclear interaction, their effective masses are equal to their restmasses, i.e.  
$U_i(\vec{r},\vec{p_i}) \equiv 0$ for mesons. 
The r.h.s. (i.e. the collision integral) 
of eq. (\ref{buueq}) describes the  
changes of $f_i(\vec{r}, \vec{p_i}, t)$ due to two-body collisions among  
the hadrons ($h_1 + h_2 \leftrightarrow h_3 + h_4$) 
and two-body decays of baryonic and mesonic resonances. Furthermore, 
$v_{12}\frac{d\sigma_{12 \rightarrow 34}}{d\Omega}$ is the in-medium collision rate, 
$\bar{f}_i = 1 - f_i\, (i = 1, .., 4)$ are the Pauli-blocking
factors for fermions and $v_{12}$ is the relative velocity between 
the colliding hadrons  
$h_1$ and $h_2$ in their center-of-mass system. In the collision integrals 
describing two-body decays of resonances one has to replace the product 
(relative velocity $\times$  cross-section $\times \ f_2$) 
by the corresponding decay rate and to introduce the proper fermion blocking
factors in the final channel. The factor $g$ in 
(\ref{buueq}) stands for the 
spin degeneracy of the particles participating in the collision 
whereas $\sum_{2,3,4}$ stands for the sum over
the isospin degrees of freedom of particles 2, 3  and 4. 
We include the following elastic and inelastic 
baryon-baryon collisions, inelastic meson-baryon collisions as 
well as meson-meson collisions (for a detailed discussion of the cross sections
employed cf. \cite{teis96}): 
\bea 
N N & \longleftrightarrow & N N \nonumber \\
N R & \longleftrightarrow & N R \non\\
N N & \longleftrightarrow & N R \non \\
N R & \longleftrightarrow & N R'\non \\
N N & \longleftrightarrow & \Delta(1232) \Delta(1232) \non \\
R & \longleftrightarrow & N \pi \non \\
R & \longleftrightarrow & N \pi \pi \non \\
R  & \longleftrightarrow & \Delta(1232) \pi,\, N(1440)\pi, \, N \rho,\, N \sigma \non \\
N(1535) & \longleftrightarrow & N \eta \non \\
N N & \longleftrightarrow & N N \pi \non  \\
\rho & \longleftrightarrow & \pi \pi \, \,\, \, \,  \mbox{(p-wave)}\non \\
\sigma & \longleftrightarrow & \pi \pi \, \, \, \, \, \mbox{(s-wave)}, \label{reacs} 
\eea
where $R$ and $R^\prime$ represent all baryonic resonances up to a mass of 1.95 GeV/c$^2$ 
\cite{teis96}. 
\end{subsection}
\begin{subsection}{The Coulomb Potential}\label{sec_coul}
Charged baryons and mesons are propagated in the electromagnetic field 
generated by the 4-current of all charged particles. Since our numerical 
studies did not indicate any sensitivity to the hadron trajectories from
the time dependent magnetic fields, we only discuss the effects of the
time-like component (i.e. $V_C$) in the following. 
The Coulomb potential in our transport model is evaluated by solving 
the static Poisson-equation in three dimensions
\bea  
 - \left( \frac{\partial^2}{\partial x^2} + \frac{\partial^2}{\partial x^2} + 
\frac{\partial^2}{\partial x^2} \right) V_C(x,y,z) = 4 \pi  \alpha \hbar c 
 \, \rho_C(x,y,z) \label{poiseq}
\eea 
($\alpha = 1/137$) in every time-step of the calculation where all 
charged particles present contribute to the charge 
density $\rho_C(x,y,z)$. The Poisson-equation (\ref{poiseq}) is integrated 
by means of the {\it Alternating-Direction Implicit Iterative} 
(ADI-)algorithm \cite{varga62,stoer80}, where the boundary conditions are obtained 
by a multipole expansion up to the quadrupole moment of the Coulomb potential. 
Since the mesons are propagated as free particles with respect to the nuclear
mean-field, the Coulomb force
\be
\vec{F}_c \left(\vec{r}\right) = - q \,
\vec{\nabla}_r V_c \left(\vec{r}\right)
\label{coul_force}
\ee
(besides the Lorentz force from the magnetic field) is the 
only force acting on a meson with charge $q$. As noted before, the curl
of the spatial components of the vector potential is found to be small in
the overlap regime of the colliding ions and can be neglected within the
numerical accuracy achieved. For charged baryons
$F_c$ represents a force in addition to the nuclear force 
from the scalar mean-field potential $U(\vec{r},\vec{p})$,
\bea 
\frac{d \vec{r}_i(t)}{d t} & = & \frac{\vec{p_i}}{E_i} + \frac{m_i^*}{E_i}\;
\vec{\nabla}_p \; U(\vec{r}_i,\, \vec{p}_i(t) ) \non \\
\frac{d \vec{p}_i(t)}{d t} & = & - \frac{m_i^*}{E_i}\;
\vec{\nabla}_r \; U(\vec{r}_i,\, \vec{p}_i(t) ) - q_i \vec{\nabla} 
V_C(\vec{r_i}), \label{tpeoms}
\eea
however, is of minor importance here. 
\end{subsection}
\end{section}
\begin{section}{Coulomb analysis of pion spectra}\label{secres}
In the following we investigate the Coulomb-effects on charged pion
spectra in heavy-ion collisions in the SIS energy range. We especially
look at the correlation between the Coulomb-forces acting on
charged pions and their freeze-out times and freeze-out densities,
respectively. Finally, we present pion spectra and $\pi^-/\pi^+$ ratios
from our transport calculation for $Au + Au$ at 1 GeV/A and $Ni + Ni$ at
2.0 GeV/A which are studied experimentally at SIS.
\begin{subsection}{Correlation between Coulomb-effects and pion
production times}
In the high density phase of a heavy-ion collision in the SIS-energy range
baryonic resonances like the $\Delta(1232)$, $N(1440)$, $N(1535)$ etc. are
excited in nucleon-nucleon collisions or via multi-step processes in
pion-nucleon or resonance-nucleon collisions. These excited baryonic
resonances then can be deexcited in collisions with nucleons 
(e.g. $\Delta(1232)N
\rightarrow NN$) or decay into nucleons and pions. If these decays take place
in the high density phase of the heavy-ion reaction the emerging pions are
likely to be absorbed again, but if the decays occur in the expansion phase of the
reaction or in low density regimes, i.e. at the surface of the reaction volume,
the pions may escape from the nuclear medium (freeze-out). 
\\
This line of argument is supported by
fig. \ref{fig_deco}
\begin{figure}
\epsfig{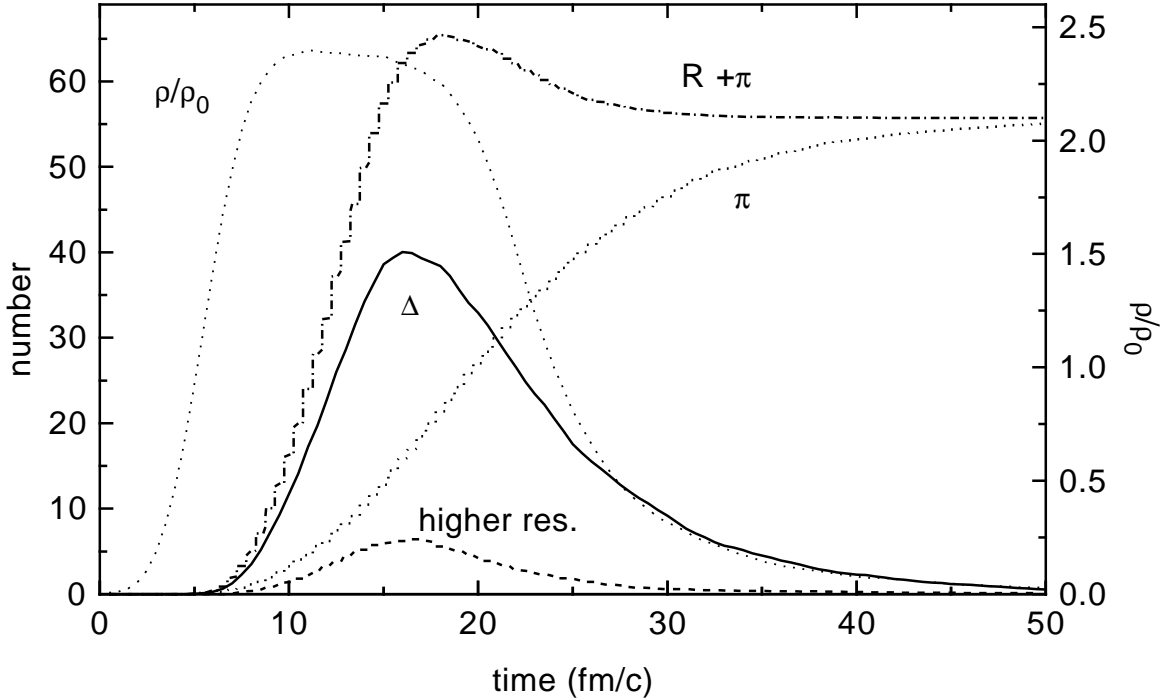}
\vspace{-4.5cm}
\caption{Number of particles in the transport calculation as a function
of time for the reaction $Au + Au$ at 1.0 GeV/A and impact parameter $b = 0$ fm:
solid line: $\Delta(1232)$; dashed line: sum of all higher resonances; 
dotted line: pions; dash-dotted line: sum of baryonic resonances and 
pions. The short-dashed line shows the baryon density in a volume
of $1 fm^{3}$ in the center of the CMS of the colliding heavy-ions.}
\label{fig_deco}
\end{figure}
where we show the number of baryonic resonances and pions present in
our transport calculation as a function of time for a central ($b = 0$ fm)
$Au + Au$ collision at 1.0 GeV/A. The short-dashed line indicates the baryon
density in a volume of $1$ $fm^{3}$ in the center of the CMS of the two
colliding nuclei. The population of the baryonic resonances sets in as the
density in the overlap zone of the colliding heavy-ions increases and reaches
its maximum at the end of the high density phase at about $15$ - $20$ $fm/c$.
This holds for the $\Delta(1232)$ (solid line) as well as for the
higher resonances (dashed and dotted lines). For the pions, however, we obtain
a different picture. The number of pions increases monotonically with time
and reaches its maximum (asymptotic) value at the end of the reaction, which is
about a factor 1.2 higher than the maximum number of $\Delta$'s in the high
density phase.
This implies that most of the pions observed at the end of the heavy-ion reaction
stem from resonances decaying in the expansion phase of the reaction. \\
In order to discuss this effect more quantitatively we show in fig.
\ref{fig_numtime} the number of asymptotically observed pions as a function of their
freeze-out times for CMS momenta of $0.2$ (solid line), $0.6$ (dashed line)
and $1.0$ GeV/c (dotted line) for a $Au + Au$ collision at 1.0 GeV/A in central
collisions ($ b = 0$ fm). 
The production of pions in all three
kinematical regions yields about $10 \%$ of its
maximum at $t \approx 15$ $fm/c$, reaches its maximum between $20$ and $30$ $fm/c$ and
then decreases with time. Since the production of pions with low momenta
requires less energy than the creation of high momentum pions, there are
quantitative differences in the production times of pions in the different
kinematical regimes\footnote{The 'production' time here is defined as 
the time when
the $\Delta$ decays into a nucleon and a free pion that propagates out to the
detector without any further strong interactions.}.
Pions with low momenta are already produced in the
beginning of the overlap of both nuclei while the production of pions with
high momenta starts only if there is enough energy density piled-up in order to
produce them. For a
similar reason the relative production rate of high momentum pions falls off
more quickly with time than that of pions with low momenta. In general,
the hard pions stem dominantly from the heavier baryonic resonances 
\cite{teis96}. 
Light resonances live on average longer than heavy resonances 
since their decay width is smaller than that of heavy resonances. 
This is particularly true for the $\Delta$ whose momentum-dependent width
leads to a rather long lifetime for low mass $\Delta$-resonances. 
A second
reason is that light resonances can be populated not only in the compressed
phase of a heavy-ion reaction but also in less energetic nucleon-nucleon
collisions in the expansion phase of the reaction while the creation of
heavy resonances - equivalent to the production of hard pions -
needs more energetic two-body collisions which are only present 
in the high density phase of
the heavy-ion collision. Due to this fact we observe that the maxima of the
pion-distributions in fig. \ref{fig_numtime} are shifted to smaller
times $t$ as the pions become harder. \\
This can be seen more quantitatively
in fig. \ref{fig_timevsp} where the average production time of pions 
produced in
central events is shown as a function of their momentum in the CMS of the two colliding
heavy-ions for the reaction $Au + Au$ at 1.0 GeV/A. The squares indicate
the result of the CBUU-calculation and the solid line corresponds to a linear
fit to the calculated results. From this figure we can infer that soft pions are
emitted on average approximately $10$ $fm/c$ later than hard pions with
momenta of about $1.0$ GeV/c.
\begin{figure}
\epsfig{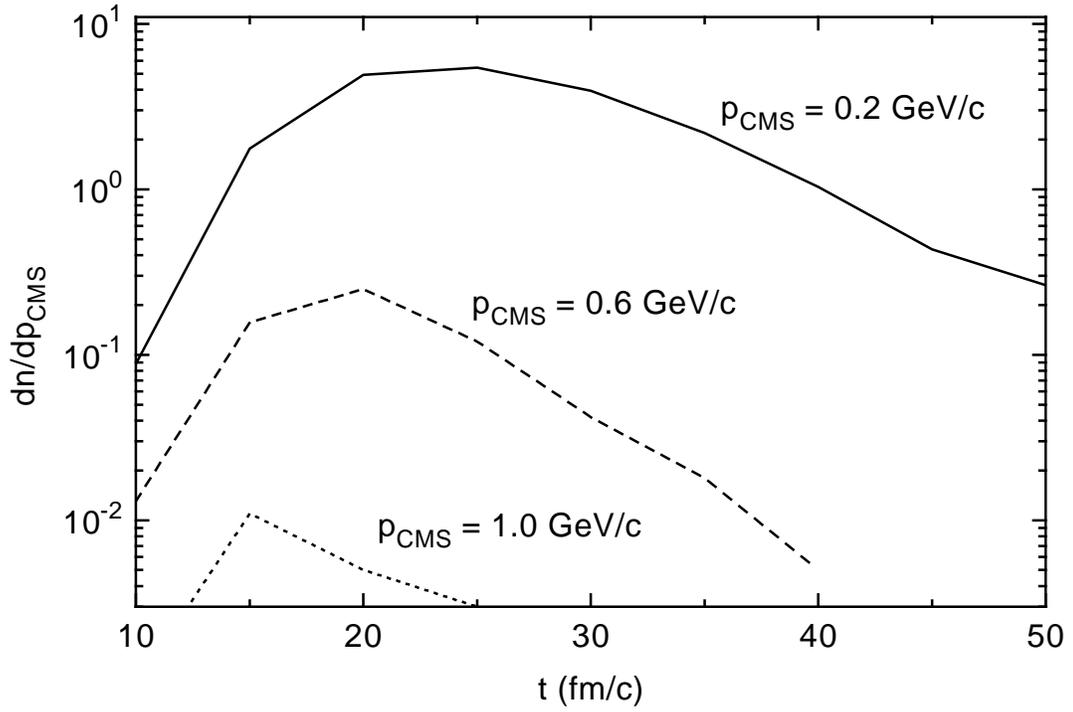}
\vspace{-4.5cm}
\caption{Number of emerging pions as a function of their final production time
for $Au + Au$ at 1.0 GeV/A taking into account only central
events ($b = 0$ fm) for different CMS pion momenta;
solid line: $p_{cms} = 0.2$ GeV/c, dashed line: $p_{cms} = 0.6$ GeV/c, 
dotted line: $p_{cms} = 1.0$ GeV/c.}
\label{fig_numtime}
\end{figure}
\begin{figure}
\epsfig{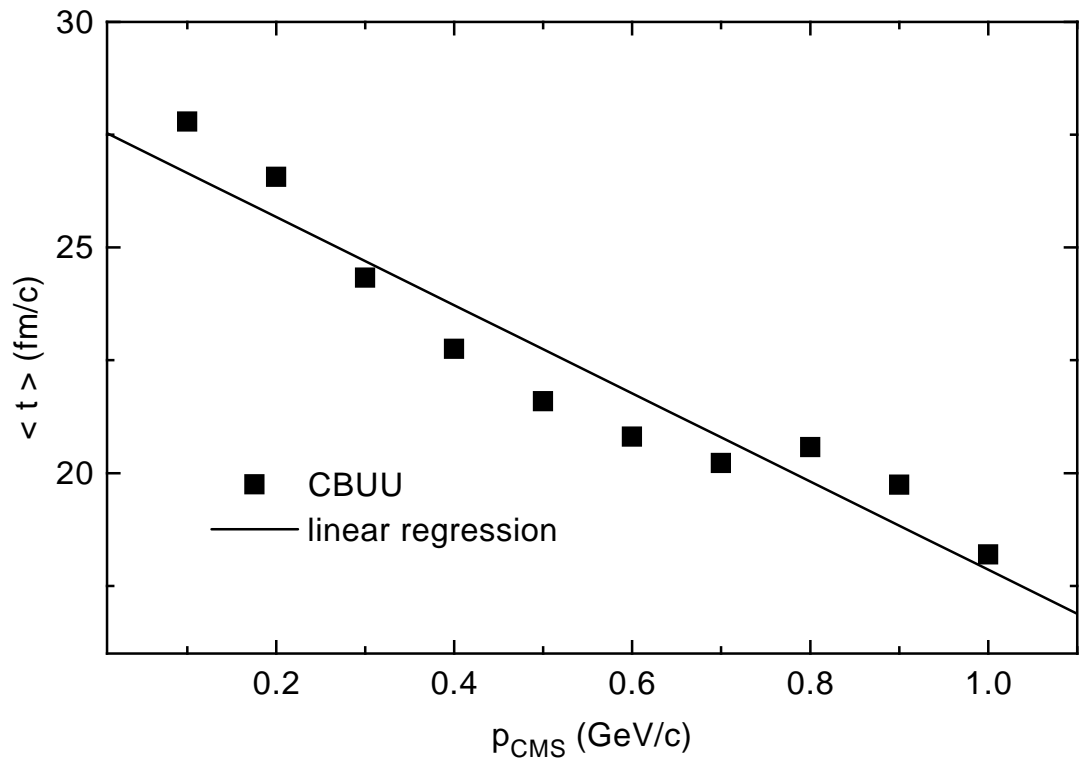}
\vspace{-4.5cm}
\caption{Average production time of pions as a function of their
CMS momentum for $Au + Au$ at 1.0 GeV/A taking into account
only central events ($b = 0$ fm); squares: results of the
CBUU-calculation, solid line: linear fit to the CBUU-results.}
\label{fig_timevsp}
\end{figure}
\begin{figure}
\epsfig{file=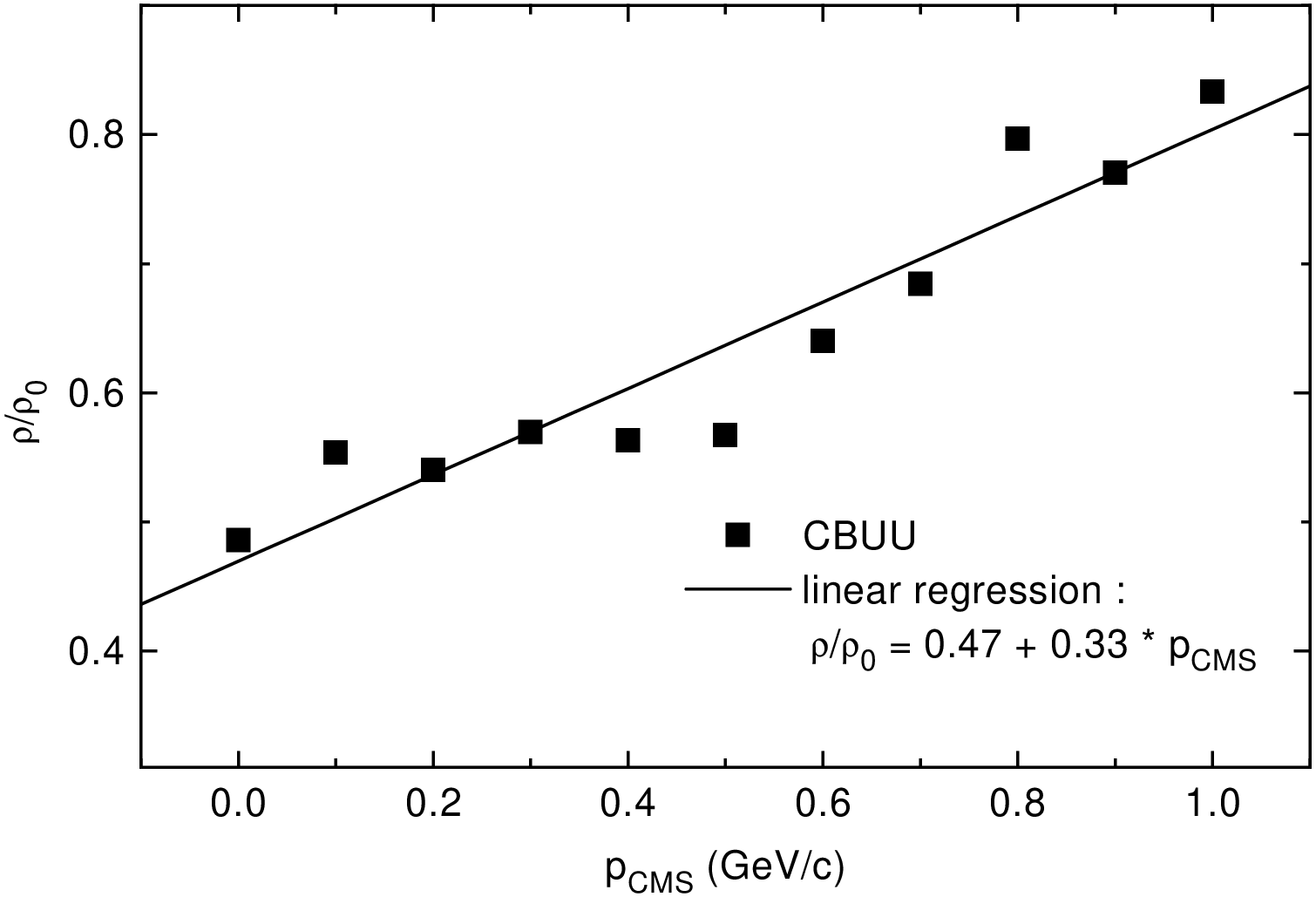,width=112mm}
\vspace{-4.5cm}
\caption{Average production density of pions as a function of their
CMS momentum for $Au + Au$ at 1.0 GeV/A taking into account
all events ($0 \le b \le 16$ fm); squares: results of the
CBUU-calculation, solid line: linear fit to the CBUU-results.}
\label{fig_densvsp}
\end{figure}
During this time the baryonic matter expands before the emission 
of the soft pions. Thus the production density for hard pions is on average
higher than that of soft pions. \\
The relation between the average production density and the
pion momentum in the CMS of the heavy-ion reaction is depicted in
fig. \ref{fig_densvsp} for the reaction $Au + Au$ at 1.0 GeV/A
taking into account all events ($0 \le b \le 16$ fm). Quantitatively we
can conclude that pions with momenta of $\approx 1.0$ GeV/c are emitted
on average at $0.8 \rho_0$. After production of the hard pions the
fireball expands to $\approx 0.5\rho_0$ before most of the soft pions are
emitted. Note, that both densities are below normal nuclear matter 
density such that pions carry information essentially only on the low
density phase of the reaction. 
From the linear dependence of the freeze-out density
on the pion momentum we can infer that the Coulomb potential acting
on the pions, when leaving the reaction zone, must
also be a linear function of the pion momenta because the charge density is
proportional to the baryon density. Indeed this behaviour can be found by means
of our transport calculation. In the top part of fig. \ref{fig_coulvsp}
the absolute magnitude of the average Coulomb potential 
felt by the pion at its production point is plotted
versus the pion momentum in the CMS. The open circles indicate the potential
of the $\pi^-$ and the open triangles that for $\pi^+$ 
while the solid squares denote
the Coulomb potential averaged over $\pi^+$ and $\pi^-$. The solid line
corresponds to a linear fit to the averaged values (squares) in the momentum
range between $0.1$ and $1.0$ GeV/c. We find that the average Coulomb
potential at freeze-out increases linearly with the pion momentum except
for pions with momenta below $50$ MeV/c. 

In the bottom part of fig.
\ref{fig_coulvsp} we investigate how the Coulomb effects on the pion
distributions change as a function of the centrality of the heavy-ion reaction:
the squares and the circles show the average Coulomb potential of the pions
 at freeze-out as a function of their CMS momentum for peripheral and central
collisions, respectively. The solid and dashed lines represent the corresponding
linear fits to the values obtained from the CBUU-calculation.
\begin{figure}
\hspace{2cm}
\epsfig{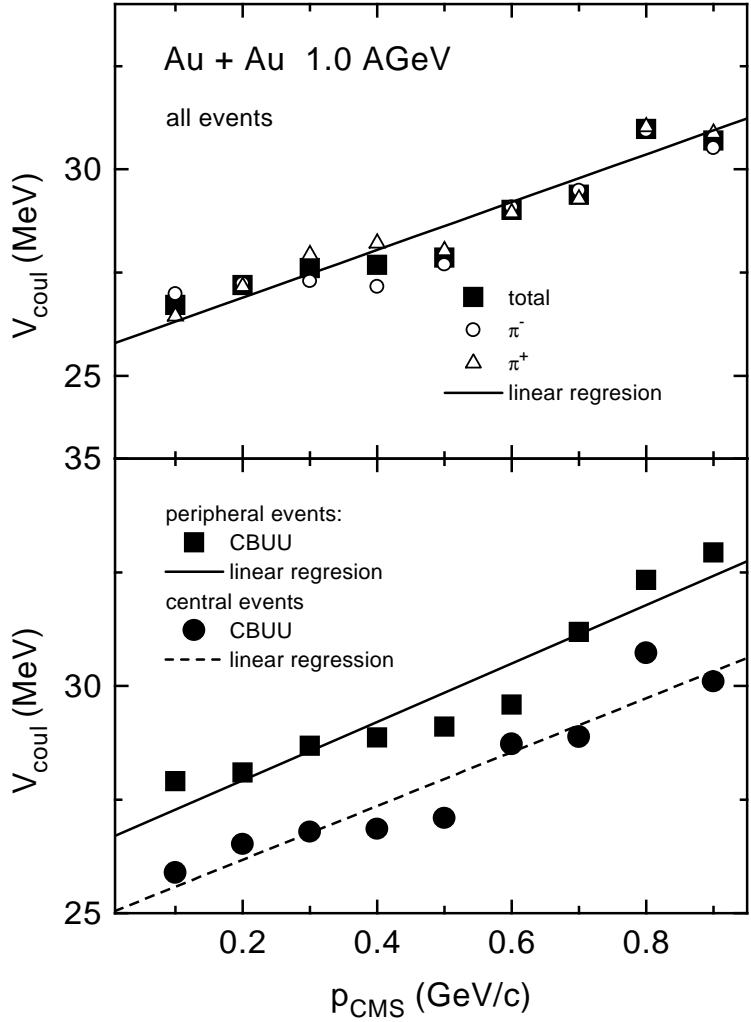}
\caption{Average Coulomb potential of pions (at their production time)
as a function of
their CMS momentum for $Au + Au$ at 1.0 GeV/A.
Top: The circles represent the results obtained for $\pi^-$ and the
triangles those for $\pi^+$. The filled squares indicate the
Coulomb potential averaged over $\pi^+$ and $\pi^-$ while the solid line
is a linear fit to these values for pion momenta between $0.1$ and
$1.0$ GeV/c. Here all events are taken into account (0 $\le$ b $\le$ 16 fm).
Bottom: The squares indicate the result for the Coulomb potential
averaged over $\pi^+$ and $\pi^-$ for peripheral collisions ($ b > 6$ fm)
and the circles represent the same for central
collisions ($0 \le b \le 6$ fm). The solid and the dashed line represent
the corresponding linear fits to the CBUU-calculations.}
\label{fig_coulvsp}
\end{figure}
Fig. \ref{fig_coulvsp} clearly indicates that the Coulomb effects on the final
pion distributions are also sensitive to the reaction geometry, i.e. pions
created in peripheral collisions feel a $\approx 15 \%$ stronger Coulomb
potential than those produced in central collisions. While
in central collisions almost all nucleons participate in the reaction, 
get compressed and collectively expand again, more
and more nucleons do not participate in the heavy-ion reaction as the
collision becomes more peripheral with increasing impact parameter. As a 
consequence the pion freeze-out configuration is more compact in peripheral
reactions than in central collisions such that the pions see a stronger
Coulomb field when decoupling from the baryons.

By establishing a correlation between the momenta of pions emitted in
heavy-ion collisions and their average production times and densities
we have proven that pions are a sensitive probe to the expansion phase
of heavy-ion reactions. Obviously, by measuring the pion
distributions in heavy-ion collisions one cannot directly deduce their
freeze-out density or their freeze-out time, but one can take advantage
of the sensitivity of the pion distributions to the Coulomb potential
due to the presence of charged baryons and mesons in the reaction
zone by investigating the differences in the observed
$\pi^+$- and $\pi^-$-spectra. This can be achieved by looking at the
ratio of the cross sections for $\pi^+$- and $\pi^-$-production. This
ratio
\bea
R = \frac{\sigma(\pi^-)}{\sigma(\pi^+)} \label{defratio}
\eea
is shown in fig. \ref{fig_ratiowo} as a function of the pion kinetic
energy in the CMS of the heavy-ion collision for the reaction $Au + Au$
at $1.0$ GeV/A. The top of fig. \ref{fig_ratiowo} shows the results obtained 
with no Coulomb effects included while the bottom part depicts $R$ 
obtained in a calculation including Coulomb effects. 
\begin{figure}
\hspace{2cm}
\epsfig{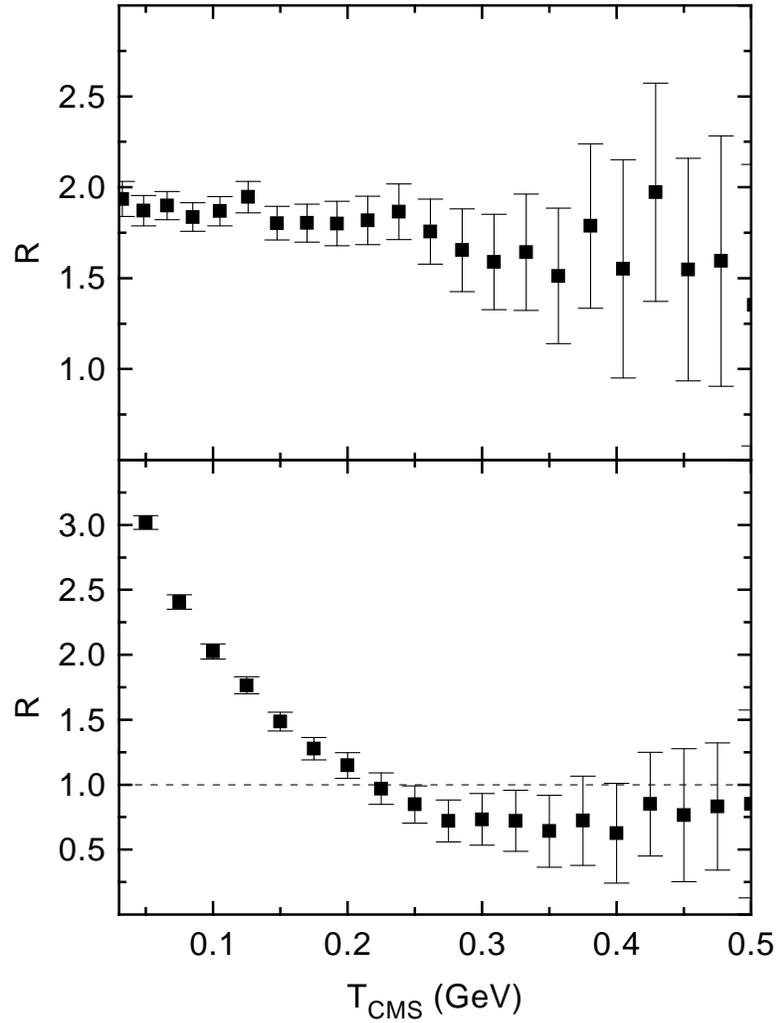}
\caption{The ratio $R$ as a function of the
CMS pion kinetic energy for $Au + Au$ at $1.0$ GeV/A obtained in a CBUU-calculation 
without Coulomb effects (top) and in a calculation including the 
Coulomb effects (bottom). The squares denote the corresponding 
computational results and 
the errorbars account for the statistical uncertainty.} 
\label{fig_ratiowo}
\end{figure}
Besides the (numerical) statistical fluctuations the ratio $R$ is
constant ($\approx 1.85 \pm 0.1$) when neglecting the Coulomb forces since 
the pions propagate freely out of the nuclear medium. Hence the $\pi^-$ and
$\pi^+$-spectra are not shifted relative to each other and the deviation of the
ratio $R$ from $1$ reflects the isospin asymmetry of the $Au + Au$ system
which according to \cite{stock86} reads,
\bea
R = \frac{5N^2 + NZ}{5Z^2 + NZ}, \label{isorat}
\eea
where $N$ and $Z$ are the number of neutrons and protons, respectively.
For a $Au + Au$ collision eq. (\ref{isorat}) gives $R = 1.95$, which is
in good agreement with 
our transport calculation without Coulomb forces.
The ratio obtained when including the
Coulomb forces (bottom part of fig. \ref{fig_ratiowo}) shows an energy-dependent behaviour. For kinetic energies
of $\approx 50$ MeV it yields a value of about $3.0$ whereas with increasing
pion kinetic
energy the $\pi^-$/$\pi^+$-ratio decreases to about $0.8$ for
energies of $\approx 300$ MeV. This energy-dependent behaviour of the ratio
$R$ can be attributed to the Coulomb potential which results in opposite
forces for $\pi^+$ and $\pi^-$ mesons and which on average has a different
strength for pions in different kinematical regions
(cf. fig. \ref{fig_coulvsp}).
\end{subsection}
\begin{subsection}{Sensitivity of the $\pi^-$/$\pi^+$-ratios to the
reaction geometry}\label{sens_geo}
In the previous section we have shown that the variation in the
Coulomb potential felt by pions due to their different production
densities is the origin of the energy-dependent $\pi^-$/$\pi^+$-ratio
$R$. In this subsection we will investigate the sensitivity of $R$ to the reaction
geometry of the heavy-ion collision. In this respect we display in fig. 
\ref{fig_geoang} the ratio $R$ for the reaction $Au + Au$
at $1.0$ GeV/A as a function of the pion CMS kinetic energy for different
CMS-angles of the outgoing pions. The solid line shows a fit to the total ratio $R$, while the dashed,
the dotted and the dash-dotted lines represent the resulting $\pi^-/\pi^+$ 
ratio for
$\Theta_{CMS} = 0^o \pm 22.5^o$, $45^o \pm 22.5^o$ and $90^o \pm 22.5^o$,
respectively. We observe a large ratio
for soft pions emitted in forward direction while the ratios
for slow pions emerging at $\Theta_{CMS} = 45^o$ or
$90^o$ are significantly smaller. This large difference in R
vanishes if the pions become harder. For pions with CMS kinetic energies
of $\approx 250$ MeV the effect has almost disappeared. This angular dependence
of the ratio $R$ can be understood within the 'participant-spectator'
picture. While in central collisions almost all target and projectile
nucleons participate in the collision and therefore form expanding
hot nuclear matter of densities below $\rho_0$, in peripheral collisions the
spectator matter provides more compact sources for the Coulomb potential
(cf.  sect. 3.1). 

The dramatic increase in the ratio $R$ for soft pions moving in forward direction
is due to the fact that these pions move in the same direction as
the spectator matter and therefore stay for a long time relatively
close to the spectators. This results in a larger Coulomb distortion of the
pion-spectra than for those pions leaving the reaction zone in other
directions. This argument is supported by fig. \ref{fig_geoimp} where in the  
top part we show the ratio $R$ for pions moving in forward direction 
($\Theta_{CMS} = 0^o \pm 22.5^o$) for $Au + Au$ at $1.0$ 
GeV/A as a function of the pion kinetic CMS-energy for different event classes.
The dotted histogram represents the resulting $R$ taking into
account only central ($0 \le b \le 6$ fm) events, the dashed histogram
corresponds to $R$ obtained for peripheral events ($7 \le b \le 16$fm),
while the solid histogram corresponds to the ratio when
including all events ($0 \le b \le 16$ fm).
\begin{figure}
\epsfig{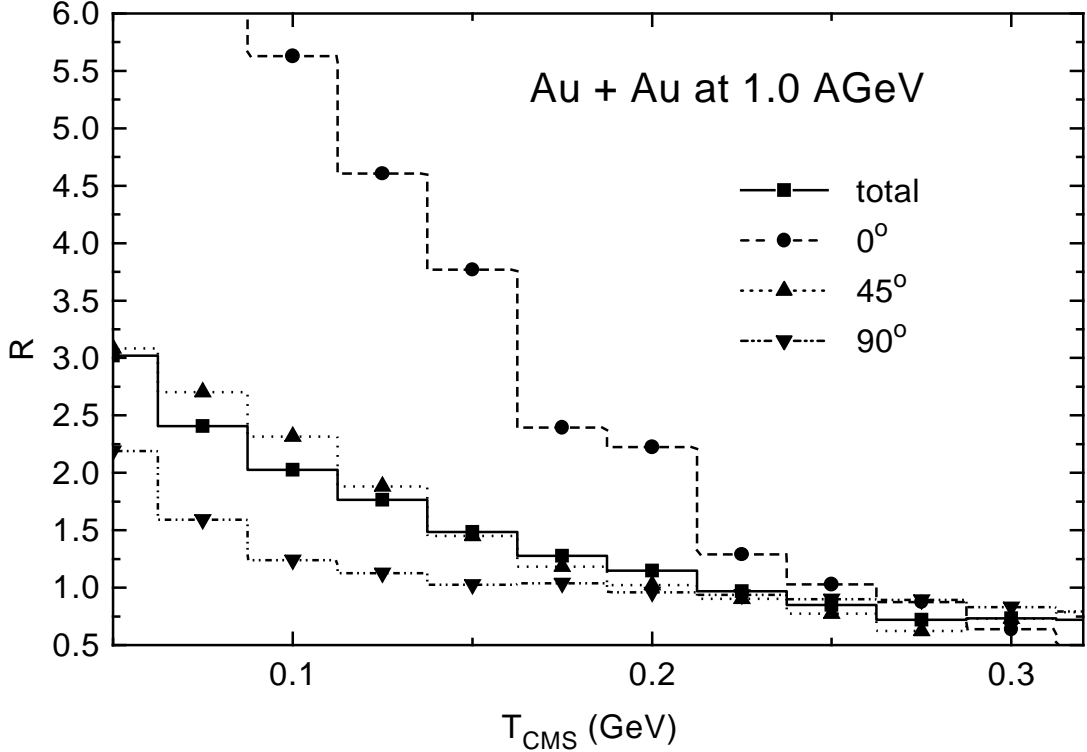}
\vspace{-4.5cm}
\caption{The ratio $R$ for the reaction $Au + Au$ at $1.0$ GeV/A as a
function of the CMS pion kinetic energy for different CMS polar
angles: $\Theta_{CMS} = 0^o \pm 22.5^o$ dashed line,
$\Theta_{CMS} = 45^o \pm 22.5^o$ dotted line,
$\Theta_{CMS} = 90^o \pm 22.5^o$ dash-dotted line. The solid line represents
the ratio $R$ integrated over all angles.}
\label{fig_geoang}
\end{figure}
\begin{figure}
\hspace{2cm}
\epsfig{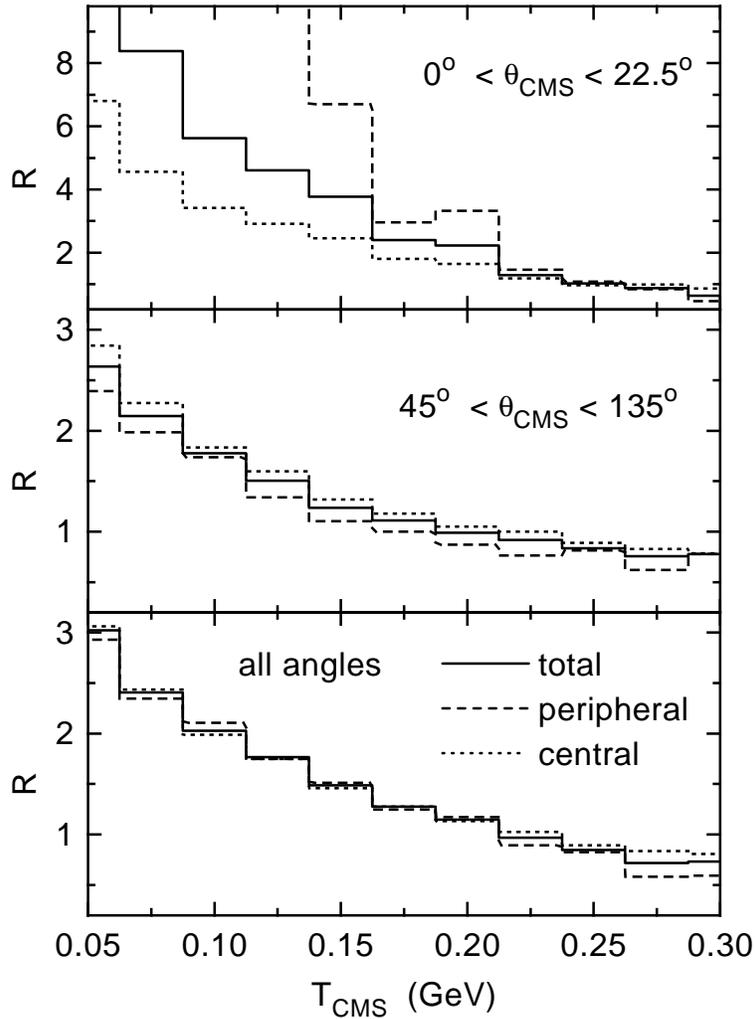}
\caption{The ratio $R$ for the reaction $Au + Au$ at $1.0$ GeV/A as a
function of the CMS pion kinetic energy for pions 
with different event classes: all events: solid histogram; central events
($0 \le b \le 6$ fm): dotted histogram; peripheral events
($7 \le b \le 16$fm): dashed histogram. Top: Pions moving in 
forward direction ($0^o \le \Theta_{CMS} \le 22.5^o$); middle: pions 
leaving the reaction zone with $45^o \le \Theta_{CMS} \le 135^o$; 
bottom: all pions ($0^o \le \Theta_{CMS} \le 180^o$).}
\label{fig_geoimp}
\end{figure}
Comparing the three histograms in the top of fig. \ref{fig_geoimp} we 
find a strong
sensitivity of $R$ to the impact parameter of the heavy-ion collision
which is directly related to the impact parameter dependence of the pion 
Coulomb potential (cf. fig. \ref{fig_coulvsp}).
For soft pions we obtain a larger value of $R$ for peripheral events
than for central events. The size of this effect is again
reduced for harder pions. This variation of $R$ with the impact parameter
of the heavy-ion collision supports the argument that the Coulomb
attraction/repulsion for pions moving in forward direction is mainly due
to the spectator matter. The larger amount of spectator matter in peripheral
collisions leads to larger Coulomb effects than in central collisions.
This dependence of the Coulomb-ratio on the impact parameter of the
heavy-ion reaction is found to be reversed for pions with kinetic energies 
below $200$ MeV when considering pions leaving the reaction zone under 
$45^o \le \Theta_{CMS} \le 135^o$ as shown in the middle of fig. 
\ref{fig_geoimp}. 
The dependence of $R$ on the impact parameter vanishes  
within the statistical uncertainties of our transport calculation
when evaluating the Coulomb ratio $R$ for all pions 
produced in the reaction (bottom part of fig. 
\ref{fig_geoimp}). 
This can 
be interpreted in the same line of argument: slow pions  
produced in peripheral collisions moving perpendicular to the direction 
of the spectators feel weaker Coulomb forces than those produced in central 
collisions because the spectator matter moves away from the reaction zone 
and thus reduces the effective Coulomb potential 'seen' by the pions. 
\end{subsection}
\begin{subsection}{Comparision with experimental data}\label{sec_compdata} 
In this subsection we compare our calculations for the inclusive pion
spectra to the respective experimental data \cite{Wagner} and 
present  $\pi^-/\pi^+$ ratios for $Au + Au$ at 1 GeV/A and Ni + Ni at 2.0 GeV/A.
 
First we compare our calculation with the inclusive experimental $\pi^+$-spectrum for 
Au +Au at 1.0 GeV/A  for lab. angles in the range $40^0 \le \Theta_{lab}
\le 48^0$ \cite{Wagner} in fig. \ref{incl}. The full squares represent the
experimental data of the KaoS collaboration as a function of the laboratory
momentum $p_{lab}$ whereas the histograms are the
result of our calculation including the uncertainty due to the numerical
statistics (shaded area). We overestimate
the experimental spectrum in the range $p_{lab} \approx$ 300-400 MeV/c for the
heavy system Au + Au, but reproduce the high momentum part of the spectrum
reasonably well within the numerical statistics achieved. This general trend is
also found in the IQMD calculations by Bass et al. \cite{Bass} for this system.
We note, that the high momentum tail of the spectrum basically stems for
higher baryon resonances as discussed in more detail in \cite{teis96}. 
\begin{figure}
\epsfig{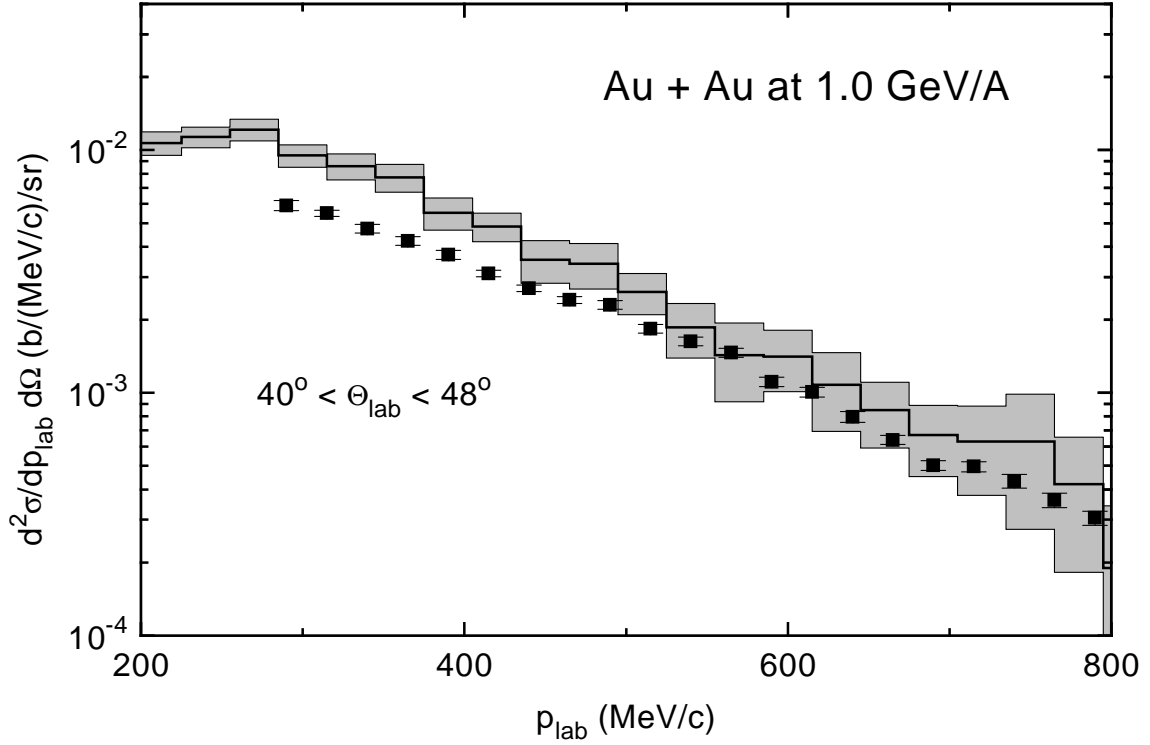}
\vspace{-4.5cm}
\caption{The inclusive $\pi^+$-spectrum for 
Au +Au at 1.0 GeV/A  for lab. angles in the range $40^0 \le \Theta_{lab}
\le 48^0$. The full squares represent the
experimental data of the KaoS collaboration \cite{Wagner} 
as a function of the laboratory
momentum $p_{lab}$ whereas the histograms are the
result of our calculation including the uncertainty due to the numerical
statistics (shaded area).}
\label{incl}
\end{figure}
\\
In fig. \ref{fig_au10rat} we show $R$ for $Au + Au$ at 1 GeV/A as a function of the  
pion kinetic energy in the CMS integrated over all angles (top), 
for $\Theta_{lab} = 45^o$ (middle) and $\Theta_{lab} = 90^o$ (bottom). 
Whereas the ratio R drops almost linearly with the pion energy up to about 0.25GeV
when integrating over all angles, its variation with the pion energy is more
pronounced for the two angular cuts. Here, especially the $\pi^-/\pi^+$ ratio at low
pion energy ($\le$ 0.1 GeV) changes significantly with laboratory angle;
its drops by about 20\% for $\Theta_{lab} \approx$ 45$^o$ and increases by
about 20\% for $\Theta_{lab} \approx$ 90$^o$ relative to the angle averaged
value. Since the relative variation in the ratio R is quite pronounced, it
should also clearly be seen in the respective experimental pion ratios. For 
comparison we also display in fig. \ref{fig_au10rat} the experimental  
$\pi^-/\pi^+$-ratio from the FOPI-collaboration \cite{fopi} (top: open circles) as well 
as the ratio obtained by the KaoS-collaboration \cite{ratio} for 
$40^o \le \Theta_{lab} \le 48^o$ (middle: open triangels). Whereas the ratio resulting 
from the transport calculation underestimates the FOPI-data in the region of $T_{CMS} \approx 75$ 
MeV it agrees within the statistical errors for higher kinetic energies. In comparison to 
the KaoS-data we find that the ratio is described quite well at $T_{CMS} = 50$ MeV but is 
underestimated for higher pion energies. 
\begin{figure}
\hspace*{2cm}
\epsfig{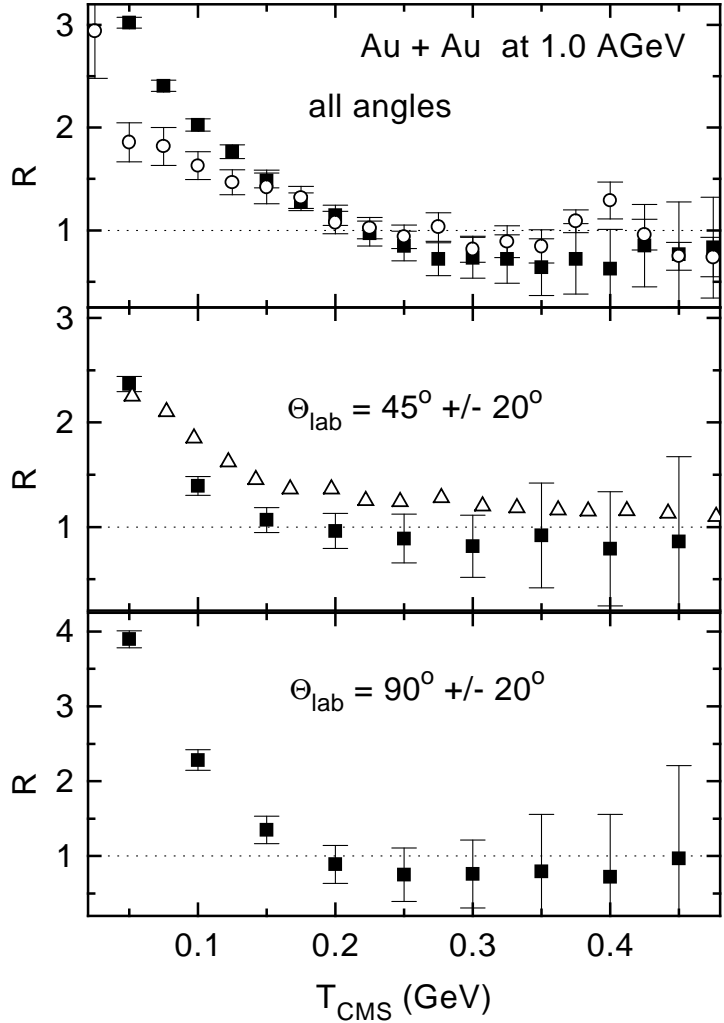}
\caption{The inclusive $\pi^-/\pi^+$ ratio R for 
Au +Au at 1.0 GeV/A  and different angular cuts in the laboratory frame
as a function of the pion kinetic energy in the CMS (squares). The errorbars indicate the
numerically accuracy within the statistics reached. The open circles represent the experimental 
data of the FOPI-collaboration \protect\cite{fopi} and the open triangles 
indicate the experimental data from the KaoS-collaboration 
\protect\cite{ratio} for $40^0 \le \Theta_{lab} \le 48^o$.}
\label{fig_au10rat}
\end{figure}
\\
In fig. \ref{fig_nirat} we present the $\pi^-/\pi^+$ ratio for Ni + Ni at 2.0 GeV/A 
for $85^o < \Theta_{CMS} < 135^o$. The solid squares represent the results of 
our transport calculation and the open circles correspond to the $\pi^-/\pi^+$-ratio 
obtained by the FOPI-collaboration for a beam energy of $1.93$ GeV/A \cite{fopi2}. Within the statistical 
errors of the calculation the data are reproduced for kinetic energies up to $\approx 500$ MeV. 
Since Coulomb effects decrease with the charge of the colliding 
system the ratio $R$ is close to $1$ when neglecting the Coulomb forces because 
$N \approx Z$. Due to the lower charge of the system the ratio 
shows only a very moderate enhancement for low pion energy.

\begin{figure}
\epsfig{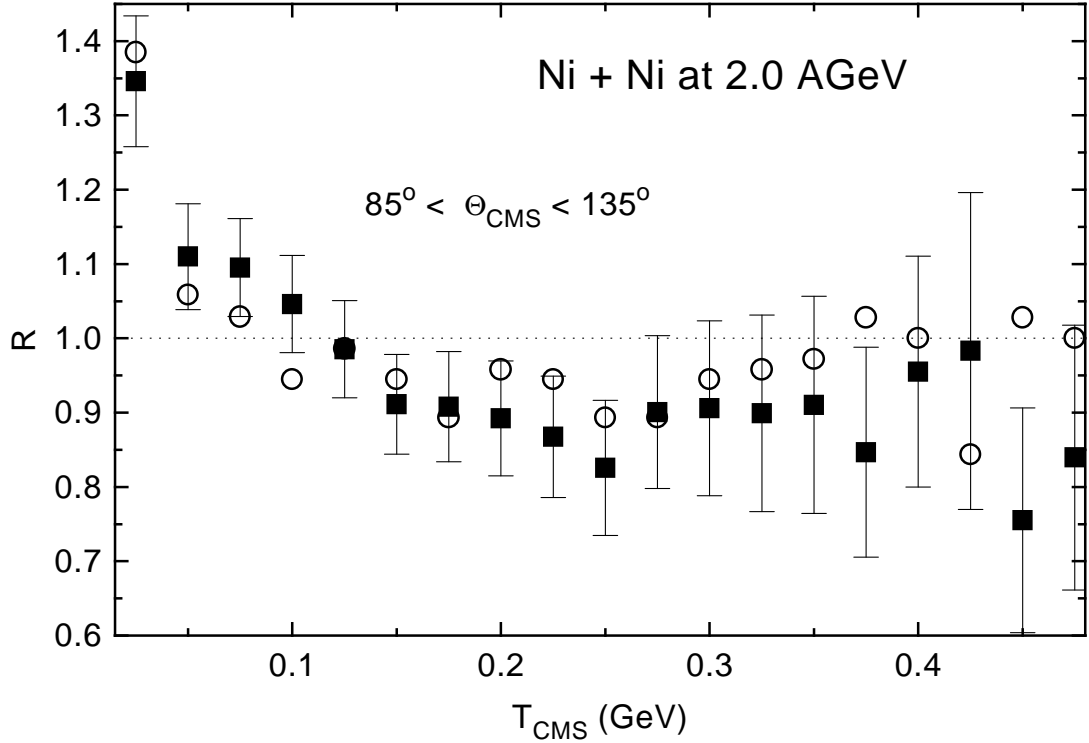}
\vspace{-4.5cm}
\caption{The inclusive $\pi^-/\pi^+$ ratio R for 
Ni + Ni at 2.0 GeV/A as a function of the pion kinetic energy in the CMS for 
$85^o < \Theta_{CMS} < 135^o$. The solid squares repesent the results of the 
CBUU-calculation; the errorbars indicate the
numerically accuracy within the statistics reached. The open circles 
represent the data obtained by the FOPI-collaboration \protect\cite{fopi2} 
for a beam energy of 1.93 GeV/A.}
\label{fig_nirat}
\end{figure}
\end{subsection}
\end{section}
\begin{section}{Summary}
In this paper we have explored the possibility to use differential $\pi^-/\pi^+$
ratios as a clock for the expansion dynamics of a heavy-ion reaction. The
analysis is performed within the coupled-channel (CBBU) transport approach
which has proven in ref. \cite{teis96} to adequately describe differential
pion spectra in the SIS energy regime. We have shown explicitly the correlation
between the pion final production time and the baryon density or Coulomb
potential that is seen by the pions at freeze-out. Furthermore, a sizeable
correlation between the pion production time and the pion momentum is found
due to the strong interaction dynamics.

The different Coulomb potentials seen by slow/fast pions in different
directions with respect to the beam axis and for different impact parameter
of the heavy-ion collisions directly induce sizeable variations in the
$\pi^-/ \pi^+$ ratios especially for low pion momenta. These ratios can
be measured directly by the FOPI and KaoS collaborations at GSI; detailed
predictions for $Au + Au$ at 1 GeV/A have
been made in this context and are awaiting further experimental control.

So far, we have performed our analysis without including any selfenergies
for the pions while assuming the mean-field potentials 
to be the same for all baryons. From the exploratory study in
ref. \cite{ehehalt} we know, that selfenergy effects show up dominantly
at low pion momenta in the CMS. However, since
such hadronic selfenergies are approximately the same
for $\pi^+$ and $\pi^-$ mesons, their effect should cancel out when considering
the ratio of the differential spectra. Precise experimental data will be
very helpful in clarifying this question.
\end{section}

\begin{section}*{Acknowledgements}
The authors like to thank A. Wagner, C. Sturm and H. Oeschler for 
clarifying discussions throughout the course of this study. 
\end{section}

\end{document}